\documentclass{aa}
\usepackage{graphicx}
\usepackage{natbib}
\usepackage{color}

\begin{document}

\title{Updated study of the quintuple system V994 Her. }

\author{P. Zasche~\inst{1} \and R. Uhla\v{r}~\inst{2} }

\offprints{Petr Zasche, \email{zasche@sirrah.troja.mff.cuni.cz}}

\institute{Astronomical Institute, Charles University in Prague, Faculty of Mathematics and Physics, CZ-180~00 Praha 8, \\
             V~Hole\v{s}ovi\v{c}k\'ach 2, Czech Republic
          \and Private Observatory, Poho\v{r}\'{\i} 71, CZ-254 01 J\'{\i}lov\'e u Prahy, Czech Republic}

\date{Received \today}

\titlerunning{The updated study of the quintuple system V994 Her.}
\authorrunning{P. Zasche \& R. Uhla\v{r}}

\abstract{}
 {We present a new updated study of the unique quintuple system of two eclipsing binaries, V994 Her.}
 {Based on new obtained photometric observations, we analysed the
period variations of the two eclipsing pairs of V994~Her.}
 {We found that the orbital period of the two eclipsing binaries around a common
barycenter is about 2.9~years instead of the original period,
which was reported as 6.3~years. Moreover, the system now
seems to be close to coplanar ($i \approx 84^\circ$), and we
also revised the individual masses of all four
eclipsing components. The two eclipsing pairs show slow apsidal motion with periods
of about 116 and 111 years, respectively. Pairs A and B are distant
by about 17~mas on the sky,
which is encouraging for observers because an interferometric detection may be possible in the next years.}
 {All of these results are based on the photometric observations of V994~Her and its detailed
analysis of the period changes of the two eclipsing binaries. }

\maketitle

\section{Introduction}

The eclipsing system \object{V994 Her} was classified as an eclipsing binary of Algol type by
\cite{1999IBVS.4659....1K}, who also published a short remark about a possible variability of the
secondary component. Even though the system was observed by the Hipparcos satellite \citep{HIP},
its true nature was not discovered until the detailed study of V994~Her by
\cite{2008MNRAS.389.1630L}. Its quadruple nature of two eclipsing binaries was recognized, which
was the first discovery of such a system in the whole sky. The system still remains interesting
today because the two eclipsing pairs are being monitored for more than 20~years now, and the
orbital evolution of both binaries can be studied.

Quadruple (or even higher-order multiple) systems like V994~Her can teach us to better
understand the formation and evolution of stars and stellar systems. They may answer questions like how such complex structures
were formed \citep{2002ARA&A..40..349T}, how they evolve into their current state
\citep{2008MNRAS.389..925T}, what the preferred initial configuration of the system parameters
is, and whether we can trace it somehow from the current configuration \citep{2005A&A...439..565G}, or we may find out the
multiplicity fraction of these stars \citep{2013ARA&A..51..269D}. All of these interesting
questions can partly be answered through a detailed study of eclipsing binaries as parts of complex
multiple systems. Following the theoretical modelling, for example, indicates that the so-called
Heggie-Hills law plays a role \citep{1975MNRAS.173..729H} for a tight close binary and a
distant companion, but also for a configuration where the mass ratio of the close pair tends to be closer to unity (see e.g.
\citealt{2007prpl.conf..133G}). These two aspects can clearly be seen in the system V994~Her. Even
the existence of planets in such a system cannot be easily ruled out. We know of several planetary
systems that orbit eclipsing binaries, for instance, Kepler 47
\citep{2012Sci...337.1511O}.

About three years ago, we published \citep{2013MNRAS.429.3472Z} the first study of the orbital motion
of the system V994~Her based on new eclipse-timing observations. We found that the
system is the only one known complex system of two eclipsing binaries orbiting a common
barycenter on a rather eccentric orbit (e=0.747) and with the orbital period of about 6.3~years.
Despite its rather poor coverage in the $O-C$ diagrams, we were able to identify the (rather sharp)
period changes of the two pairs around a periastron. From this solution it followed that the two pairs
orbit each other on a slightly inclined orbit with an inclination of about 37$^\circ$.

\section{Inconsistency of our original solution for V994~Her}

The double eclipsing system V994~Her was first correctly separated into two independent periods in
2008 by \cite{2008MNRAS.389.1630L}. The authors found that the orbital periods of the two pairs are
of about $P_A$ = 2.08 days, and $P_B$ = 1.42 days. Using detailed spectroscopic observations, they
were also able to derive the individual masses of the eclipsing components and their spectral
types: pair A (B8V + A0V) and pair B (A2V + A4V). All four components are apparently clearly
detached and are still placed on the main sequence. Another more distant (about 1$^{\prime\prime}$)
component was discovered and is listed in the  Washington Double Star Catalog \citep{WDS}, which
means that this is a quintuple stellar system. At the time of Lee et al. publication, it was the
only known such system with two eclipsing binaries located on the sky.  The number of such systems
has increased today, mainly as a result of the OGLE detections of the Magellanic Cloud eclipsing
binaries. For example, \citet{2013AcA....63..323P} listed 15 new potential double eclipsing systems
in the SMC. In addition, \cite{2012A&A...544L...3C} and \cite{2013A&A...549A..86L} announced the
discovery of new interesting double eclipsing systems. However, a similar study of the period
changes for their new discovered systems is still difficult because we lack data spanning a longer
time interval.

However, our previous result on the system published in \cite{2013MNRAS.429.3472Z} seems to be
incorrect. The inclination of the eclipsing orbits and their mutual orbit of about 37$^\circ$ is an
improbable geometrical configuration and not very stable on longer periods of time. Because of the
Kozai cycles (see e.g. \citealt{2006Ap&SS.304...75E}), the two orbits should tend to be coplanar or
to have a different inclination. And finally, the new minima time observations also significantly
deviated from the published predicted fit of V994~Her (for both pairs).

\begin{figure}
  \centering
  \includegraphics[width=85mm]{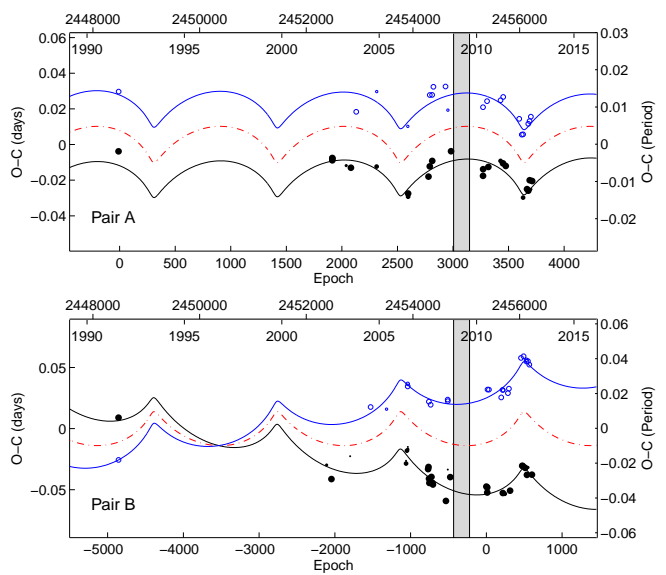}
  \caption{Plot of the $O-C$ diagrams of the two pairs A and B. The dots stand for the primary, while the
  open circles for the secondary minima, the bigger the symbol, the higher the weight. The red
  dash-dotted lines indicate the LTTE fit, while the black and blue curves represent the final
  fit (LTTE plus the apsidal motion). This solution is taken from our original study in 2013; 
    the grey area indicates where observations are missing exactly in
  between the two periastron passages.}
  \label{FigOC1}
\end{figure}

From these reasons we decided to continue monitoring this interesting eclipsing system in
the years after our study in 2013 was published. Since then, 36 new observations of minima (of both pairs
A and B) were obtained by the authors in seasons 2013, 2014, and 2015  during more than
980 days. The data used for the analysis are presented in Table \ref{MINIMA}. From studying these
new data points, we found that our original hypothesis was incorrect because we had insufficient data.
This is well visible in the plot in Fig.\ref{FigOC1}, where we plot $O-C$ diagrams for
pairs A and B according to our original solution together with the grey area that represents the time
epoch without observations. Unfortunately, this epoch of lacking data is placed exactly in
between the two periastron passages in 2006 and 2012. Hence, this paper should not be considered as
an erratum, but rather as a reanalysis of the system with new data (new times of minima), as
well as a reanalysis of the older radial velocities by \cite{2008MNRAS.389.1630L}. We finally
introduce the correct hypothesis of the true system configuration.

\section{Updated analysis}

The star with the new data was analysed using the same approach as in our previous study. This means
that the times of minima were fitted using all 16 parameters for the orbits of the two binaries (the two
ephemerides, two apsidal motion hypotheses, and the period changes that are due to the orbit around a
common barycenter). The well-known light-travel time effect (LTTE) hypothesis \citep{Irwin1959}
was used to describe the variation in the $O-C$ diagrams.

\begin{figure}
  \centering
  \includegraphics[width=85mm]{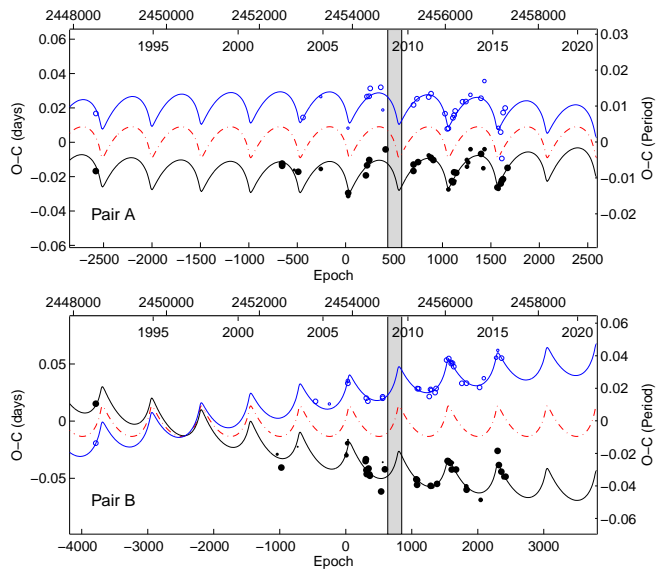}
  \caption{Updated plot of the $O-C$ diagrams of the two pairs A and B (see Fig.\ref{FigOC1} for
  description). The missed periastron passage in 2009 is clearly seen in the grey area.}
  \label{FigOC2}
\end{figure}

In Fig. \ref{FigOC2} we plot the updated $O-C$ diagrams for the
two A and B pairs, showing their
period changes (the y-axis) with respect to time (the x-axis). Our current
hypothesis describes the observed data much better (the $\chi^2$ resulted in a value about  five times lower). The parameters describing the LTTE fit are given below in Table \ref{Tab_param}. The
orbital period of the two pairs resulted in about half the value presented in our previous study
\citep{2013MNRAS.429.3472Z}, which is also due to the missing data near the periastron passage in
2009. With the new data the sharp curvatures near the periastron passages are clearly visible and
well covered with the observations.

\begin{figure}
  \centering
  \includegraphics[width=85mm]{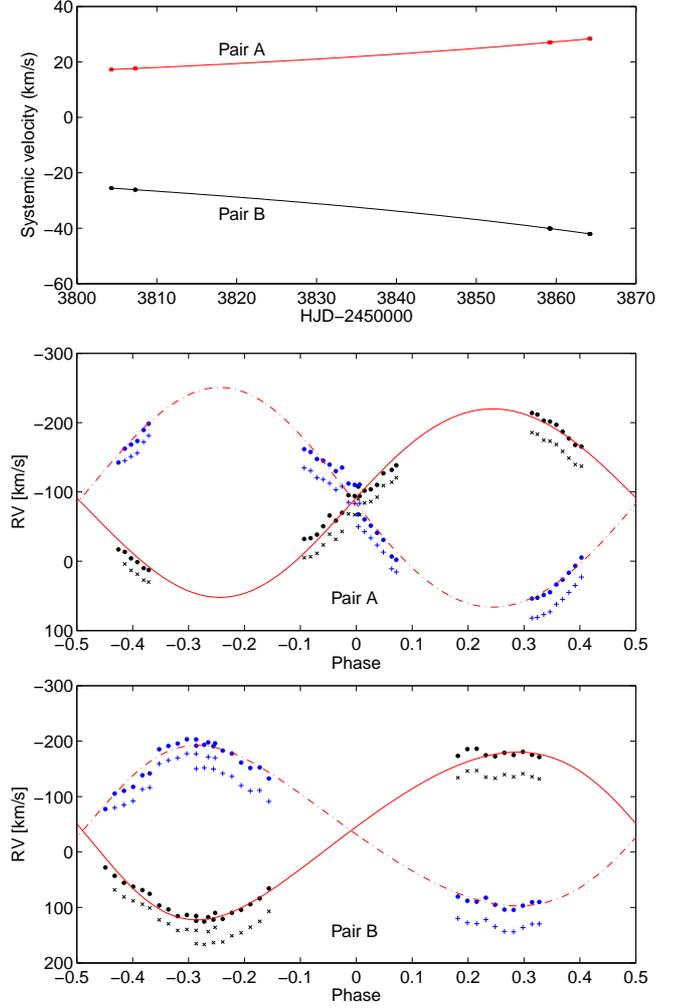}
  \caption{Radial velocities of V994 Her as re-interpreted from the original paper by \cite{2008MNRAS.389.1630L}.
  \emph{Top}: Radial velocity corrections of the two A and B pairs as these pairs revolve on the orbit around
  the common barycenter. \emph{Middle}: Radial velocities of pair A (before (plus and cross signs) and
  after (dot signs) the correction). \emph{Bottom}: Radial velocities of pair B (the same notation).}
  \label{FigRVs}
\end{figure}

Our hypothesis based on the much shorter orbital period of the two pairs is also able to solve the
problem of radial velocities published in \cite{2008MNRAS.389.1630L}. The radial
velocities they reported are somewhat distorted. Especially theose of pair B were unusual because some jumps between the individual datasets near the quadrature are visible. Using our approach, we are
also able to compute the radial velocities on the mutual orbit and subtract this contribution from
the radial velocities. This is shown in Figure \ref{FigRVs}, where the calculated radial velocities
are shown exactly in the time epochs when the data by \cite{2008MNRAS.389.1630L} were obtained. The difference between the first and the last data point of about $10~km \cdot s^{-1}$
for pair A and of about $16~km \cdot s^{-1}$ for pair B is significant enough for such a shift.
Hence, we applied this correction to the radial velocities by \cite{2008MNRAS.389.1630L} and
reanalysed their data. We found the mass ratios to be only slightly different (the change
is below 0.05). We discuss the masses in the next section.

\section{Results}

\begin{table}
 \tiny
 \scalebox{0.9}{
 \begin{minipage}{90mm}
  \caption{New heliocentric minima of V994 Her used for the analysis.} \label{MINIMA}
   \tiny
  \begin{tabular}{@{}l l l l l l@{}}
\hline
HJD - 2400000 & Error & Pair & Type  & Filter & Observer/Ref.\\
 \hline
56368.61655 & 0.00029 &   A  &  sec  &    I   &  RU - this paper \\
56441.53093 & 0.00020 &   A  &  sec  &    R   &  RU - this paper \\
56463.37125 & 0.00081 &   A  &  prim &    C   &  RU - this paper \\
56465.45059 & 0.00259 &   A  &  prim &    C   &  RU - this paper \\
56766.52219 & 0.00090 &   A  &  sec  &    R   &  RU - this paper \\
56767.53147 & 0.00027 &   A  &  prim &    R   &  RU - this paper \\
56839.44646 & 0.00247 &   A  &  sec  &    C   &  RU - this paper \\
56842.53174 & 0.00132 &   A  &  prim &    C   &  RU - this paper \\
57119.58372 & 0.00049 &   A  &  prim &    C   &  RU - this paper \\
57141.49265 & 0.00127 &   A  &  sec  &    R   &  RU - this paper \\
57142.49880 & 0.00135 &   A  &  prim &    R   &  RU - this paper \\
57189.40527 & 0.00119 &   A  &  sec  &    R   &  RU - this paper \\
57190.41699 & 0.00066 &   A  &  prim &    R   &  RU - this paper \\
57214.38908 & 0.00093 &   A  &  sec  &    R   &  RU - this paper \\
57215.41792 & 0.00039 &   A  &  prim &    R   &  RU - this paper \\
57240.41791 & 0.00029 &   A  &  prim &    R   &  RU - this paper \\
57241.49825 & 0.00055 &   A  &  sec  &    R   &  RU - this paper \\
57287.33273 & 0.00062 &   A  &  sec  &    C   &  RU - this paper \\
57336.25467 & 0.00038 &   A  &  prim &    C   &  RU - this paper \\  \hline

56354.60974 & 0.00207 &   B  &  sec  &    C   &  RU - this paper \\
56450.37231 & 0.00204 &   B  &  prim &    C   &  RU - this paper \\
56455.43271 & 0.00040 &   B  &  sec  &    C   &  RU - this paper \\
56457.46946 & 0.00054 &   B  &  prim &    R   &  RU - this paper \\
56736.59690 & 0.00049 &   B  &  sec  &    R   &  RU - this paper \\
56758.50910 & 0.00259 &   B  &  prim &    R   &  RU - this paper \\
56827.48758 & 0.00078 &   B  &  sec  &    I   &  RU - this paper \\
57091.63291 & 0.00116 &   B  &  sec  &    R   &  RU - this paper \\
57123.50228 & 0.00057 &   B  &  prim &    R   &  RU - this paper \\
57128.56049 & 0.00279 &   B  &  sec  &    R   &  RU - this paper \\
57133.42997 & 0.00248 &   B  &  prim &    R   &  RU - this paper \\
57150.47067 & 0.00049 &   B  &  prim &    R   &  RU - this paper \\
57204.42643 & 0.00054 &   B  &  prim &    R   &  RU - this paper \\
57209.49597 & 0.00062 &   B  &  sec  &    I   &  RU - this paper \\
57241.34185 & 0.00229 &   B  &  prim &    R   &  RU - this paper \\
57248.44349 & 0.00149 &   B  &  prim &    R   &  RU - this paper \\
57295.30457 & 0.00069 &   B  &  prim &    R   &  RU - this paper \\ \hline
  \hline
 \end{tabular}
\end{minipage}}
\end{table}

\begin{table*}
 \centering
  \caption{Updated orbital parameters for V994~Her.}  \label{Tab_param}
  \begin{tabular}{c c | c c | c c}
\hline
 & & \multicolumn{2}{c}{Pair A} & \multicolumn{2}{c}{Pair B} \\
 Parameter & Unit & Value & Error & Value  & Error \\
 \hline
 $JD_0$ &  HJD  & 2453855.1339 & 0.0034    & 2453857.4361 & 0.0054 \\
 $P$    &  Day  & 2.0832649    & 0.0000029 &  1.4200401   &  0.0000042 \\
 $e$    &       &   0.0307     &   0.0085  &   0.1253     &  0.0069   \\
 $\omega$& Deg  &     9.9      &    6.0    &   300.7      &  4.9   \\
 $\mathrm{d}\omega/\mathrm{d}t$ & Deg/Cycle & 0.0018 & 0.0012  & 0.0126 & 0.0073 \\
 $p_3$  & Year  &    2.91      &   0.14    &    2.91      &  0.14 \\
 $T_0$  &  HJD  & 2456045      &   122     &  2456045     &  122\\
 $A$    & Day   &  0.0091      &   0.0032  &  0.0134      &  0.0037 \\
 $\omega_3$ & Deg &  236.2     &  26.7     &   56.2       &  26.7 \\
 $e_3$  &       &   0.758      &   0.071   &   0.758      &  0.071 \\
 \hline
\end{tabular}
\end{table*}

Our hypothesis led to several important results, which are completely different from the findings
published in the previous analysis. At first, the orbital motion has a period of only about 2.9 years. The other parameters also
slightly changed, but the amplitudes and relatively high
eccentricity of the mutual orbit remained almost unchanged. However, the apsidal motion of the two
pairs resulted in $(116 \pm 50)$ years for pair A and $(111 \pm 40)$ for pair B. But the most
significant change is the resulting geometry of the whole system, which now seems to be close to
coplanar (i.e. the orbits of the two eclipsing pairs and  their mutual orbit seem to lie very
close to one plane, see below). And moreover, if we adopt the Hipparcos \citep{HIP} value of the
parallax as correct ($\pi_{HIP} = (3.90 \pm 0.74)$~mas), the
predicted angular separation of the components is about $\Delta \alpha_{12} = (16.7 \pm
5.6)$~mas (instead of the original $\Delta \alpha_{12} = (27.6 \pm 6.8)$ mas). This value is
still rather high and should encourage a prospective interferometric direct detection of the two pairs
in the sky. However, the attempt of detection should be made in specific phases on
the 2.9 yr orbit to be successful.

\begin{figure}
  \centering
  \includegraphics[width=85mm]{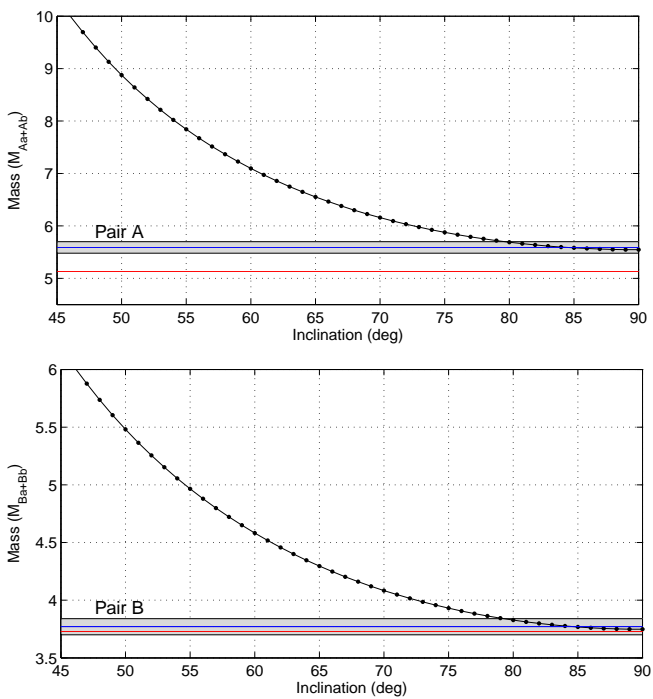}
  \caption{Plot of the masses (total mass of the pair A and B) as derived from the analysis of period
  changes with respect to the inclination of the 2.9 yr orbit. The red solid lines represent the masses
  reported by \cite{2008MNRAS.389.1630L}, the blue lines stand for the masses from our analysis (with the grey
  area showing the uncertainties). The black line represents the different masses with respect to
  the inclination between the orbits.}
  \label{FigMass}
\end{figure}

As a byproduct of our analysis, we also found that the masses of all components as derived by
\cite{2008MNRAS.389.1630L} need to be reconsidered. The masses of the two pairs were calculated
using the fact that a so-called mass function of the third body (see e.g. \citealt{Mayer1990}) can
be computed from the orbital parameters, and the inclinations of the two orbits are
also close to 90$^\circ$, that is,
$$\frac{m_A \cdot A_A}{\sin i_A} = \frac{m_B \cdot A_B}{\sin i_B} \Longrightarrow m_A \cdot A_A
\approx m_B \cdot A_B \doteq 0.050,$$ where the masses of pairs are denoted as $m_A$, and $m_B$ (in
solar masses), respectively, and the amplitudes of the period variations in the $O-C$ diagrams for
the two pairs are labelled $A_A$ and $A_B$ (in days). The correct mass values are slightly
higher for pair A but remained similar for pair B, see Table \ref{Tab_masses} for
a comparison of \cite{2008MNRAS.389.1630L} and our results. The plot with the resulting masses for the A and B binaries with respect to the inclination is plotted in Fig. \ref{FigMass}. The derived mass of pair A is significantly higher than the value
published by \cite{2008MNRAS.389.1630L}. Moreover, the plot shows that the inclination of
the wide orbit is probably close to 90$^\circ$, hence the system is close to coplanar. This means that the
mutual inclination angle is $i \approx (84 \pm 10)^\circ$ for pair A and $i \approx
(85 \pm 12)^\circ$ for pair B.

\begin{table}
 \centering
  \caption{V994~Her: Updated component masses.}  \label{Tab_masses}
  \begin{tabular}{c c c }
\hline
                      & \cite{2008MNRAS.389.1630L} & This paper\\
 \hline
 $M_{Aa}$ [M$_\odot$] & 2.83 $\pm$ 0.20 &  3.01 $\pm$ 0.06 \\
 $M_{Ab}$ [M$_\odot$] & 2.30 $\pm$ 0.16 &  2.58 $\pm$ 0.05 \\
 $M_{Ba}$ [M$_\odot$] & 1.87 $\pm$ 0.12 &  1.84 $\pm$ 0.03 \\
 $M_{Bb}$ [M$_\odot$] & 1.86 $\pm$ 0.12 &  1.93 $\pm$ 0.04 \\
 \hline
\end{tabular}
\end{table}

The question also arises whether it is possible to detect some orbital precession when the two orbits are
close to coplanar, and whether we can see some mutual eclipses of all components in the future. Only
a dedicated spectroscopic, photometric, and interferometric monitoring is able to answer this
question.

\section{Conclusion}

The system V994~Her was reanalysed with new photometric observations. Our solution yielded a very
different geometrical configuration than originally assumed. In this 2+2 quadruple system with
coplanar orbits, the precession of all orbits is only very slow and hard to detect. This system
structure is much more probable and seems to be adequately stable for a long period of time. There
still remain some open questions, such as the other parameters of the 2.9-year orbit, and the
confirmation that the distant component also belongs to the system (which would make this a
quintuplet). However, the most promising way to proceed today seems to be an interferometric
detection of the two components because they should have a semi-major axis of about 17~mas and a
magnitude difference between the components of about 1.4~mag. This is well within the limits for
modern stellar interferometers.

The main advantage of the system V994~Her is that it is bright enough for photometric monitoring by
only small telescopes (all of our new data were obtained with telescopes with apertures of 20 cm or
smaller). It is also quite promising that the periods are relatively short, which means that this
is  an ideal laboratory of celestial mechanics.

\section*{Acknowledgments}
We would like to thank David Vokrouhlick\'y for a useful discussion and valuable comments. We
also thank the anonymous referee for helpful and critical suggestions. This work was
supported by the Czech Science Foundation grants no. P209/10/0715 and GA15-02112S. This research
has made use of the Washington Double Star Catalog maintained at the U.S. Naval Observatory, the
SIMBAD database, operated at CDS, Strasbourg, France, and of NASA's Astrophysics Data System
Bibliographic Services.


\begin{thebibliography}{99}
 \bibitem[\protect\citeauthoryear{Caga{\v s} \& Pejcha}{2012}]{2012A&A...544L...3C} Caga{\v s}, P., \& Pejcha, O.\ 2012, A\&A, 544, L3
 \bibitem[\protect\citeauthoryear{Duch{\^e}ne \& Kraus}{2013}]{2013ARA&A..51..269D} Duch{\^e}ne, G., \& Kraus, A.\ 2013, ARA\&A, 51, 269
 \bibitem[\protect\citeauthoryear{Eggleton \& Kisseleva-Eggleton}{2006}]{2006Ap&SS.304...75E} Eggleton, P.~P., \& Kisseleva-Eggleton, L.\ 2006, Ap\&SS, 304, 75
 \bibitem[\protect\citeauthoryear{Goodwin \& Kroupa}{2005}]{2005A&A...439..565G} Goodwin, S.~P., \& Kroupa, P.\ 2005, A\&A, 439, 565
 \bibitem[\protect\citeauthoryear{Goodwin et al.}{2007}]{2007prpl.conf..133G} Goodwin, S.~P., Kroupa, P., Goodman, A., \& Burkert, A.\ 2007, Protostars and Planets V, 133
 \bibitem[\protect\citeauthoryear{Heggie}{1975}]{1975MNRAS.173..729H} Heggie, D.~C.\ 1975, MNRAS, 173, 729
 \bibitem[\protect\citeauthoryear{Irwin}{1959}]{Irwin1959} {Irwin}, J.~B. 1959, AJ, 64, 149
 \bibitem[\protect\citeauthoryear{Kazarovets et al.}{1999}]{1999IBVS.4659....1K} Kazarovets, E.~V., Samus, N.~N., Durlevich, O.~V., et al.\ 1999, IBVS, 4659, 1
 \bibitem[\protect\citeauthoryear{Lee et al.}{2008}]{2008MNRAS.389.1630L} Lee, C.-U., Kim, S.-L., Lee, J.~W., et al.\ 2008, MNRAS, 389, 1630
 \bibitem[\protect\citeauthoryear{Lohr et al.}{2013}]{2013A&A...549A..86L} Lohr, M.~E., Norton, A.~J., Kolb, U.~C., et al.\ 2013, A\&A, 549, A86
 \bibitem[\protect\citeauthoryear{Mason et~al.}{2001}]{WDS} Mason, B.~D., Wycoff, G.~L., Hartkopf, W.~I., Douglass, G.~G., \& Worley, C.~E. 2001, AJ, 122, 3466
 \bibitem[\protect\citeauthoryear{Mayer}{1990}]{Mayer1990} Mayer, P.\ 1990, BAICz, 41, 231
 \bibitem[\protect\citeauthoryear{Orosz et al.}{2012}]{2012Sci...337.1511O} Orosz, J.~A., Welsh, W.~F., Carter, J.~A., et al.\ 2012, Science, 337, 1511
 \bibitem[\protect\citeauthoryear{Pawlak et al.}{2013}]{2013AcA....63..323P} Pawlak, M., Graczyk, D., Soszy{\'n}ski, I., et al.\ 2013, AcA, 63, 323
 \bibitem[\protect\citeauthoryear{Perryman et al.}{1997}]{HIP} Perryman, M.~A.~C., Lindegren, L., Kovalevsky, J., et al. 1997, A\&A, 323, L49
 \bibitem[\protect\citeauthoryear{Tohline}{2002}]{2002ARA&A..40..349T} Tohline, J.~E.\ 2002, \araa, 40, 349
 \bibitem[\protect\citeauthoryear{Tokovinin}{2008}]{2008MNRAS.389..925T} Tokovinin, A.\ 2008, MNRAS, 389, 925
 \bibitem[\protect\citeauthoryear{Zasche \& Uhla{\v r}}{2013}]{2013MNRAS.429.3472Z} Zasche, P., \& Uhla{\v r}, R.\ 2013, MNRAS, 429, 3472
\end{thebibliography}
\end{document}